\documentclass[twocolumn,aps,prl,superscriptaddress]{revtex4}

\usepackage{sidecap}
\usepackage{ulem}
\usepackage{epsfig}
\usepackage{amsmath,amssymb,amsthm}
\usepackage{graphicx}
\usepackage{bm}
\usepackage{color,soul}

\DeclareMathAlphabet{\mathpzc}{OT1}{pzc}{m}{it}

\begin{document}

\renewcommand{\textfraction}{0.00}


\newcommand{\vAi}{{\cal A}_{i_1\cdots i_n}}
\newcommand{\vAim}{{\cal A}_{i_1\cdots i_{n-1}}}
\newcommand{\vAbi}{\bar{\cal A}^{i_1\cdots i_n}}
\newcommand{\vAbim}{\bar{\cal A}^{i_1\cdots i_{n-1}}}
\newcommand{\htS}{\hat{S}}
\newcommand{\htR}{\hat{R}}
\newcommand{\htB}{\hat{B}}
\newcommand{\htD}{\hat{D}}
\newcommand{\htV}{\hat{V}}
\newcommand{\cT}{{\cal T}}
\newcommand{\cM}{{\cal M}}
\newcommand{\cMs}{{\cal M}^*}
\newcommand{\vk}{\vec{\mathbf{k}}}
\newcommand{\bk}{\bm{k}}
\newcommand{\kt}{\bm{k}_\perp}
\newcommand{\kp}{k_\perp}
\newcommand{\km}{k_\mathrm{max}}
\newcommand{\vl}{\vec{\mathbf{l}}}
\newcommand{\bl}{\bm{l}}
\newcommand{\bK}{\bm{K}}
\newcommand{\bb}{\bm{b}}
\newcommand{\qm}{q_\mathrm{max}}
\newcommand{\vp}{\vec{\mathbf{p}}}
\newcommand{\bp}{\bm{p}}
\newcommand{\vq}{\vec{\mathbf{q}}}
\newcommand{\bq}{\bm{q}}
\newcommand{\qt}{\bm{q}_\perp}
\newcommand{\qp}{q_\perp}
\newcommand{\bQ}{\bm{Q}}
\newcommand{\vx}{\vec{\mathbf{x}}}
\newcommand{\bx}{\bm{x}}
\newcommand{\tr}{{{\rm Tr\,}}}
\newcommand{\bc}{\textcolor{blue}}

\newcommand{\beq}{\begin{equation}}
\newcommand{\eeq}[1]{\label{#1} \end{equation}}
\newcommand{\ee}{\end{equation}}
\newcommand{\bea}{\begin{eqnarray}}
\newcommand{\eea}{\end{eqnarray}}
\newcommand{\beqar}{\begin{eqnarray}}
\newcommand{\eeqar}[1]{\label{#1}\end{eqnarray}}

\newcommand{\half}{{\textstyle\frac{1}{2}}}
\newcommand{\ben}{\begin{enumerate}}
\newcommand{\een}{\end{enumerate}}
\newcommand{\bit}{\begin{itemize}}
\newcommand{\eit}{\end{itemize}}
\newcommand{\ec}{\end{center}}
\newcommand{\bra}[1]{\langle {#1}|}
\newcommand{\ket}[1]{|{#1}\rangle}
\newcommand{\norm}[2]{\langle{#1}|{#2}\rangle}
\newcommand{\brac}[3]{\langle{#1}|{#2}|{#3}\rangle}
\newcommand{\hilb}{{\cal H}}
\newcommand{\pleft}{\stackrel{\leftarrow}{\partial}}
\newcommand{\pright}{\stackrel{\rightarrow}{\partial}}

\title{How to test path-length dependence in energy loss mechanisms: analysis leading to a new observable}

\author{Magdalena Djordjevic\footnote{E-mail: magda@ipb.ac.rs}}
\affiliation{Institute of Physics Belgrade, University of Belgrade, Serbia}

\author{Dusan Zigic}
\affiliation{Institute of Physics Belgrade, University of Belgrade, Serbia}

\author{Marko Djordjevic}
\affiliation{Faculty of Biology, University of Belgrade, Serbia}

\author{Jussi Auvinen}
\affiliation{Institute of Physics Belgrade, University of Belgrade, Serbia}

\begin{abstract}

When traversing QCD medium, high $p_\perp$ partons lose energy, which is typically measured by suppression, and also predicted by various energy loss models. A crucial test of different energy loss mechanisms is how the energy loss depends on the length of traversed medium (so-called path-length dependence). The upcoming experimental results will allow to, at least in principle, for the first time, clearly observe how the energy loss changes with the size of the medium, in particular, by comparing already available $Pb+Pb$ measurements with now upcoming $Xe+Xe$ data at the LHC. However, in practice, to actually perform such test, it becomes crucial to chose an optimal observable. With respect to this, a ratio of observed suppression for the two systems may seem a natural (and frequently mentioned) choice. We, however, show that extracting the path-length dependence from this observable would not be possible. We here provide an analytical derivation based on simple scaling arguments, as well as detailed numerical calculations based on our advanced energy loss framework, showing that a different observable is suitable for this purpose. We call this observable path-length sensitive suppression ratio ($R_L^{AB}$) and provide our predictions before experimental data become available. This predictions also clearly show that this observable will allow a simple comparison of the related theoretical models with the experimental data, and consequently to distinguish between different (underlying) energy loss mechanisms, which is in turn crucial for understanding properties of created QCD medium.

\end{abstract}

\maketitle

{\it Introduction:} Understanding properties of Quark-Gluon plasma (QGP)~\cite{Collins,Baym} created at LHC and RHIC experiments is a major goal of ultra-relativistic heavy ion physics~\cite{QGP1,QGP2,QGP3}, which would alow understanding properties of this form of matter at its most basic level. Energy loss of high $p_\perp$ partons traversing this medium, is an excellent probe of its properties~\cite{Bjorken}, which provided a crucial contribution~\cite{QGP1} to establishing that QGP is created at this experiments. Comparing predictions of different energy loss models, and consequently different underlying energy loss mechanisms, with experimental data, is therefore crucial for understanding properties of created QGP. However, there remains a question of how to provide the most direct comparison of the theoretical energy loss predictions with experimental data.

A most basic signature, distinguishing different energy loss models, is how the predicted energy loss depends on the length of the traversed QCD medium, which is so-called path-length dependence. This path-length dependence directly relates to different underlying energy loss mechanisms, such as radiative vs. collisional energy loss, considering optically thin vs. thick QCD medium, inclusion of finite size effects, etc. From a practical point, accurate path-length dependence also determines how well a model can explain angular average ($R_{AA}$) and angular differential ($v_2$) suppression measurements~\cite{Betz}, and is therefore directly related with a wide range of open questions in comparing predictions with experimental data, such as $v_2$ puzzle~\cite{v2Puzzle} (i.e. inability of many models to jointly explain $R_{AA}$ and $v_2$ measurements). Reasonable agreement of energy loss models with experimental data is also a necessary step to a complete approach towards QGP tomography, which requires constraining QGP properties from the point of both high and low $p_\perp$ data. Therefore, the path-length dependence is crucial for differentiating between different energy loss mechanisms, where such differentiation is necessary for investigating QCD matter created at the LHC and RHIC.

\begin{figure*}
\epsfig{file=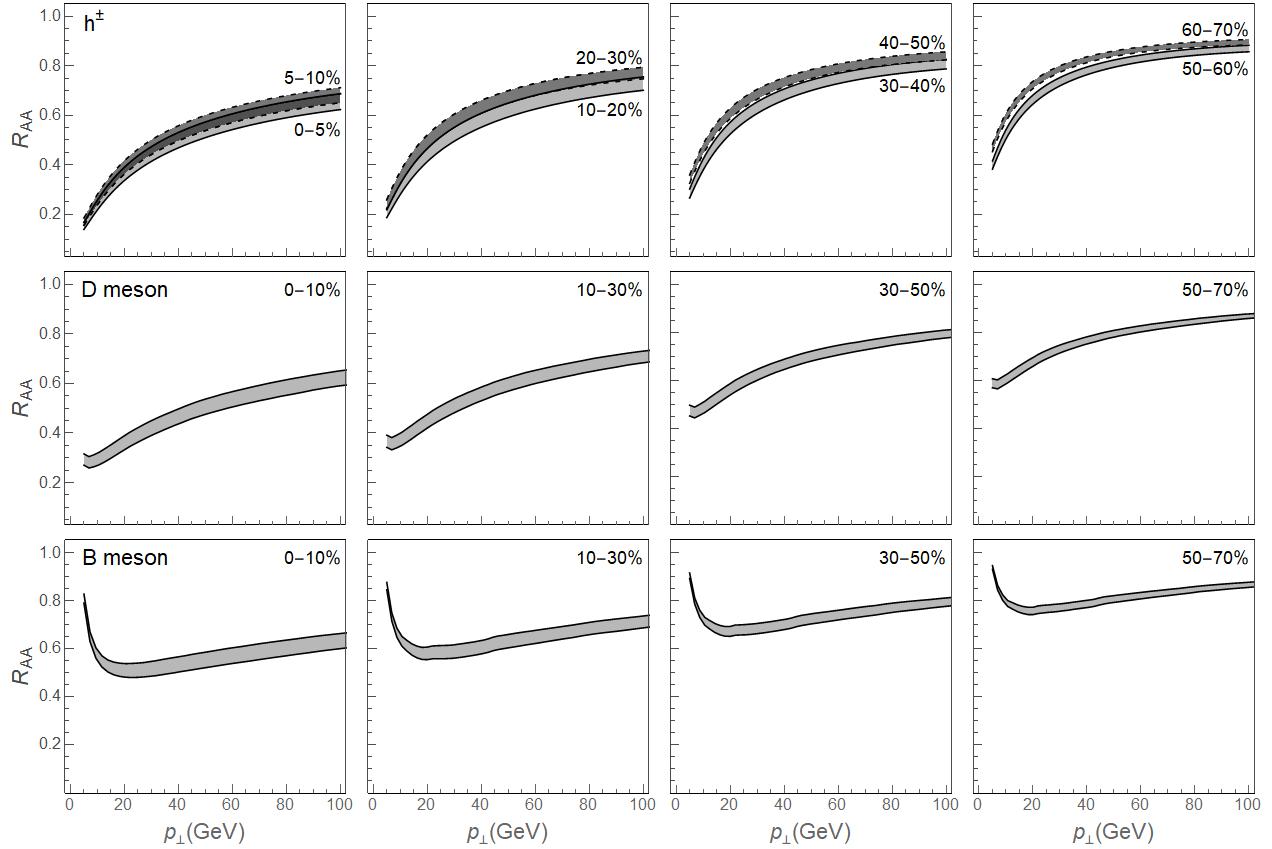,width=5.7in,height=4.2in,clip=5,angle=0}
\vspace*{-0.3cm}
\caption{ {\bf $R_{AA}$ predictions for light and heavy probes.} Theoretical predictions for light and heavy flavor $R_{AA}$ {\it vs.} $p_\perp$ in $5.44$~TeV $Xe+Xe$ collisions are shown for different centralities. Upper panels correspond to charged hadrons, middle panels to D mesons and lower panels to B mesons. For charged hadrons, full and dashed curves correspond, respectively, to predictions for  lower and higher centrality regions, where specific centralities are marked in each panel. In each panel, the upper (lower) boundary of each gray band corresponds to $\mu_M/\mu_E =0.6$ ($\mu_M/\mu_E =0.4$). }
\label{RAA}
\end{figure*}
It is intuitively clear that the most direct probe of the path-length dependence would involve comparing experimental measurements (and the related theoretical predictions) for two collision systems of different size. It is moreover intuitively clear that it would be optimal if size would be the only property distinguishing these two systems, i.e. that other properties such as average medium temperature, initial $p_\perp$ distributions, and ideally shape of the path-length probability distributions would be the same between these two systems. With respect to this, measurements for $5.02$~TeV $Pb+Pb$ collisions at the LHC are available, while measurements for $5.44$~TeV $Xe+Xe$ collisions will become available soon. As these two systems have different sizes (atomic number is $A=208$ for $Pb$, while $A=129$ for $Xe$) and the two relevant collision energies are similar, these two systems appear to make good candidates for the discussed  path-length dependence study. Moreover, we below start by showing that the other properties (stated above) of these two systems are in fact very similar, making these collisions  ideal for this study.

\begin{figure*}
\epsfig{file=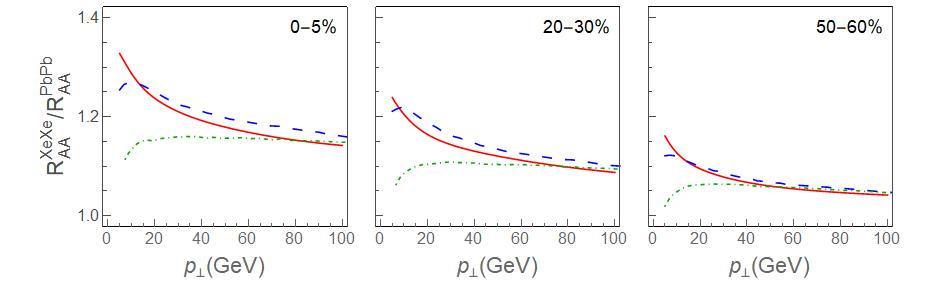,width=5.5in,height=1.9in,clip=5,angle=0}
\vspace*{-0.3cm}
\caption{ {\bf Ratio of $R_{XeXe}$ and $R_{PbPb}$ for light and heavy probes.} Theoretical predictions for the ratio of $R_{XeXe}$ and $R_{PbPb}$ as a function of $p_\perp$ is shown for charged hadrons (full curves), D mesons (dashed curves) and B mesons (dot-dashed curves). First to third panel correspond to, respectively, $0-5\%$, $20-30\%$ and $50-60\%$ centrality regions. $\mu_M/\mu_E=0.4$. }
\label{FigRAAratio}
\end{figure*}
{\it Results and Discussion: }We here consider average medium temperature, since we will below concentrate on $R_{AA}$, for which it was previously shown (see e.g.~\cite{DREENAc}) that it only weakly depends on the medium evolution. Regarding average temperature, it depends only on the ratio of the charged multiplicity and the number of participants for the particular system and centrality~\cite{DDB_PLB}. For each centrality, both of these quantities change in the two collision systems. However, it happens that, for each centrality, their ratio, and consequently the average temperature, remains the same for both collision systems. Here, our estimates of charged multiplicity and the number of participants come, respectively, from~\cite{Eskola} and~\cite{Loizides:2017ack}. Regarding initial high $p_\perp$, we previously showed~\cite{MD_5TeV} that, when the collision energy is changed almost two times (from $2.76$ to $5.02$~TeV), the influence of the change of $p_\perp$ distributions leads to only a small change (less than 10\%) in the resulting  suppression. Consequently, for the increase of less than 10\% in the collision energy (from $5.02$ to $5.44$~TeV), the same $high$ $p_\perp$ distributions can be assumed. Finally, we also calculated path-length distributions for $Xe+Xe$, at different centralities, in the same manner as previously for $Pb+Pb$~\cite{DREENAc}. It is straightforward to see that the two distributions are the same up to a rescaling factor corresponding to $A^{1/3}$~\footnote{In fact, $Xe$ is somewhat deformed~\cite{XeDef}, which can slightly influence most-central collisions (as we checked), but which does not influence more periferal collisions, which will be considered here.}. Consequently, we argue that the two systems are in fact close to ideal, when it comes to probing the path-length dependencies.

However, in addition to suitable systems, it is also crucial to choose the most suitable variable for testing path-length dependencies in different models. Since $R_{AA}$ directly increases when the system size decreases, it may seem that the ratio od $R_{AA}$ for the two systems may be a natural choice, as also frequently mentioned by experimental colleagues. To test this proposal, we start by Figure~\ref{RAA}, which provide our predictions for light (charged hadrons) and heavy (D and B meson) flavor $R_{AA}$ for $5.44$~TeV $Xe+Xe$ collisions at different centralities - these predictions can also be compared with the upcoming experimental data. These predictions are generated by our recently developed DREENA-C framework, with this framework described in~\cite{DREENAc} (all the parameters and the procedure being specified there). The framework is based on out state-of-the-art dynamical energy loss formalism~\cite{MD_Dyn}, and was previously used to obtain equivalent $R_{AA}$ predictions for $5.02$~TeV Pb+Pb collisions at the LHC~\cite{DREENAc}, which showed a good comparison with the existing data, providing confidence that the framework can reasonably explain the experimental measurements~\cite{Raa_Dyn,MD_5TeV,DDB_PLB}. Consequently, in Fig.~\ref{FigRAAratio} we show momentum dependence of $R_{AA}$ ratio for the two systems. From this figure, it would be very hard to extract the path-length dependence. For example, note that for high $p_\perp$ this ratio approaches 1, from which one may naively conclude that the underlying model has no (or only weak) path-length dependence. However, our dynamical energy loss model in fact has a strong (between linear and quadratic) path-length dependence. The same problem would emerge if experimental data would be plotted in that way, i.e. one may naively conclude that high $p_\perp$ suppression does not depend on the system size.

We next provide an analytical derivation, using simple scaling arguments, for why this problem emerges. We start from fractional energy loss $\Delta E/E$, which can be estimated as~\cite{DREENAc}
\begin{eqnarray}
\Delta E/E \sim \eta T^a L^b,
\label{ElossEstimate} \end{eqnarray}
where $a, b$ are proportionality factors, $T$ is the average temperature of the medium, $L$ is the average path-length traversed by the jet and $\eta$ is a proportionality factor (which depend on initial jet $p_\perp$). $b \rightarrow 1$ corresponds to the linear dependence, while $b \rightarrow 2$ corresponds to the quadratic (LPM like) dependence of the energy loss.

If $\Delta E/E$ is small (i.e. for higher $p_\perp$ of the initial jet, and for higher centralities), we obtain~\cite{DREENAc}
\begin{eqnarray}
R_{AA} \approx (1-\xi T^a L^b),
\label{RaaEstimate} \end{eqnarray}
where $\xi = (n-2) \eta/2$, and $n$ is the steepness of the initial momentum distribution function.

The ratio of $R_{XeXe}$ and $R_{PbPb}$ then becomes
\begin{eqnarray}
\frac{R_{XeXe}}{R_{PbPb}} &\approx& \frac{1-\xi T^a L_{Xe}^b}{1-\xi T^a L_{Pb}^b} \nonumber \\
&\approx& 1-\xi T^a L_{Pb}^b \left(1-\left(\frac{A_{Xe}}{A_{Pb}}\right)^{b/3}\right).
\label{RaaXePbRatio} 
\end{eqnarray}

One can see that this quantity is rather complicated, depending explicitly on the initial jet energy (through $\xi$), average medium temperature, and average size of the medium. Also, it explicitly depends on centrality (through $T$ and $L_{Pb}$, which decrease with increasing centrality), as seen in Fig.~\ref{FigRAAratio}. Furthermore, as centrality and initial energy of the jet increases, $\xi$, $T$ and $L_{Pb}$ become smaller, explaining why the ratio in Fig.~\ref{FigRAAratio} goes to 1 for high $p_\perp$ and high centrality, which then makes a problem of concealing the path-length dependence. That is, the ratio of $R_{AA}$s for different collision systems is not an appropriate observable for extracting path-length dependence, and a more suitable observable should reflect the coefficient $b$ in the simplest possible manner.

\begin{figure}
\epsfig{file=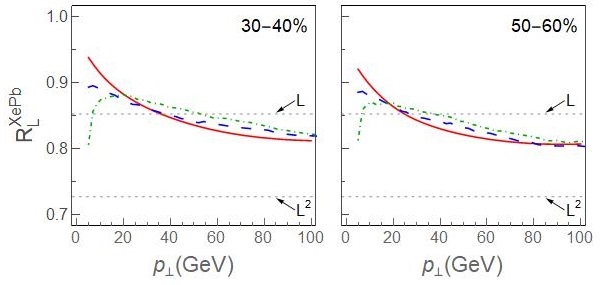,width=3.3in,height=1.7in,clip=5,angle=0}
\vspace*{-0.4cm}
\caption{ {\bf Path-length sensitive suppression ratio $R_L^{XePb}$ for light and heavy probes.} Theoretical predictions for $R_L^{XePb}$ as a function of $p_\perp$ is shown for charged hadrons (full curves), D mesons (dashed curves) and B mesons (dot-dashed curves). First and second panel correspond to, respectively, $30-40\%$ and $50-60\%$ centrality regions. $\mu_M/\mu_E=0.4$. }
\label{RLXePb}
\end{figure}

Actually, a more suitable observable $R_L^{AB}$, which we will further call path-length sensitive suppression ratio,  can be simply obtained from Eq.~\ref{RaaEstimate} by subtracting $R_{AA}$s from 1, reducing in the case of $Xe$ and $Pb$ to:
\begin{eqnarray}
R_L^{XePb} \equiv \frac{1-R_{XeXe}}{1-R_{PbPb}} \approx \frac{\xi T^a L_{Xe}^b}{\xi T^a L_{Pb}^b} \approx \left(\frac{A_{Xe}}{A_{Pb}}\right)^{b/3}.
\label{EqRLXePb} \end{eqnarray}

We see that this new quantity $R_L^{XePb}$ has a very simple form, which depends only on the medium size (through $A_{Xe}/A_{Pb}$), and on the path length dependence (through $b$). Again, note that this simple dependence holds for higher centralities and higher initial $p_\perp$, since in these regions Eq.~\ref{RaaEstimate} is applicable, leading to Eq.~\ref{EqRLXePb}. Consequently, as one plots  $R_L^{XePb}$ at higher centrality regions, one may expect that this value will approach a limit that directly reflects the path-length dependence. For example, for linear path-length dependence (i.e. $b=1$),  $R_L^{XePb}\approx 0.85$, while for quadratic (LPM like) path-length dependence (i.e. $b=2$),  $R_L^{XePb}\approx 0.73$.

Next, in Fig.~\ref{RLXePb}, we also numerically test this proposal, using DREENA-C framework, for both light and heavy flavor. Note that in contrast, we see that  $R_L^{XePb}$ is almost independent of centrality, which is exactly what one needs for such observable. Starting from higher $p_\perp>30$~GeV, where this observable is expected to work (see above), we predict a clear separation between heavy and light probes, where B meson is closest to linear, while deviation from linear dependence is more pronounced for D meson and most for charged hadrons. Finally, at high $p_\perp \rightarrow 100$~GeV, we clearly see that $R_L^{XePb}$ for all types of particles reaches a limiting value, as expected. Moreover, this limiting value ($R_L^{XePb}$=0.8) directly reflects the underlying path-length dependence, which is in our case (the dynamical energy loss formalism, with radiative and collisional energy loss in a finite size QCD medium) between linear and quadratic (i.e. $b=1.4$). Note that this extracted path-length dependence is different from a common assumption of heavy flavor having linear, while light flavor having quadratic (LPM-like) dependance. Therefore, it is clear that making this plot from experimental data, and comparing those with corresponding theoretical dependencies, can be used to differentiate between different energy loss models in a simple and direct manner. Also, note that, in distinction to Fig.~\ref{RLXePb}, where the gray dotted lines are simple and intuitive (allowing straightforward inference of path-length dependence), defining such dotted lines in Fig.~\ref{FigRAAratio} would not be possible.

{\it Summary:} We here argued that soon to come $5.44$~TeV $Xe+Xe$ experimental data at the LHC, provide previously unprecedented opportunity to distinguish between different energy loss mechanisms, and consequently for understanding properties of created QGP. This particularly when combined with expected much higher accuracy of the upcoming data. We here used both analytical and numerical analysis to propose a new observable, which is optimal for assessing the path-length dependence of the energy loss. Additionally, and independently from the proposed observable, the $R_{AA}$ predictions generated here through our state-of-the art DREENA-C formalism, can be directly compared with the upcoming $Xe+Xe$ experimental measurements. Based on our numerical results, one can expect that the path-length dependence can be straightforwardly inferred from experimental data, and compared with corresponding theoretical models. Differentiating between different energy loss mechanism is in turn crucial for high precision QGP tomography, which is future major goal of relativistic heavy ion physics.

{\em Acknowledgments:}
This work is supported by the European Research Council, grant ERC-2016-COG: 725741, and by  and by the Ministry of Science and Technological
Development of the Republic of Serbia, under project numbers ON171004 and ON173052.

\end{document}